\documentclass[prb,reprint,amsmath,amssymb,showpacs,superscriptaddress,twocolumn]{revtex4-1}

\usepackage[dvipdfmx]{graphicx}
\usepackage{bm}
\usepackage{mathrsfs}
\usepackage{color}

\DeclareMathOperator{\sgn}{sgn}
\DeclareMathOperator{\pf}{Pf}

\begin{document}
\title{Double Majorana vortex zero modes in superconducting topological crystalline insulators with surface rotation anomaly
}
\author {Shingo Kobayashi}
\affiliation{RIKEN Center for Emergent Matter Science, Wako, Saitama, 351-0198, Japan}
\author{Akira Furusaki}
\affiliation{RIKEN Center for Emergent Matter Science, Wako, Saitama, 351-0198, Japan}
\affiliation{Condensed Matter Theory Laboratory, RIKEN, Wako, Saitama, 351-0198, Japan}

\date{\today}

\begin{abstract}
The interplay of time-reversal and $n$-fold rotation symmetries ($n=2,4,6$) is known to bring a new class of topological crystalline insulators (TCIs) having $n$ surface Dirac cones due to surface rotation anomaly.
We show that the proximity-induced $s$-wave superconductivity on the surface of these TCIs yields a topological superconducting phase in which two Majorana zero modes are bound to a vortex, and that $n$-fold rotation symmetry ($n=2,4,6$) enriches the topological classification of a superconducting vortex from $\mathbb{Z}_2$ to $\mathbb{Z}_2\times\mathbb{Z}_2$.
Using a model of a three-dimensional high-spin topological insulator with $s$-wave superconductivity and two-fold rotation symmetry, we show that, with increasing chemical potential, the number of Majorana zero modes at one end of a vortex changes as $2\to1\to0$ through two topological vortex phase transitions.
In addition, we show that additional magnetic-mirror symmetry further enhances the topological classification to $\mathbb{Z} \times \mathbb{Z}$.
\end{abstract}
\maketitle

\paragraph*{Introduction.--} 
Majorana fermions bound to a superconducting vortex~\cite{Kopnin1991,Volovik1999fermion,ReadGreen2000,Volovik03,SatoFujimoto09,Teo10,SatoFujimoto16,Chiu16,Teo17review} have received great attention in recent years, since these particles obeying non-Abelian statistics~\cite{Ivanov01,Stern04,Teo10prl} were predicted to be experimentally accessible in hybrid systems of three-dimensional  (3D) topological insulators (TIs) and conventional $s$-wave superconductors (SCs)~\cite{Sato03,FuKane08}. Such non-Abelian vortices have been expected as a potential platform for topological qubits and quantum computation~\cite{Nayak08}. Recent experiments have reported evidence for Majorana fermions localized at vortex cores in superconducting TIs Bi$_2$Te$_3$~\cite{JPXu15,Sun16} and iron-based SCs~\cite{DWang18,PZhang18,QLiu18,Kong19,Machida19,Deng20}.
These topological superconducting phases have the advantage of utilizing more conventional $s$-wave pairing than intrinsic topological superconductivity mediated by odd-parity pairings~\cite{Hor10,Fu10,Sato10,Sasaki11,Sasaki12,Hashimoto15,Kobayashi15PRL,Hashimoto16,Aggarwal16,Wang16,Oudah16,Kawakami18}.   

Multiple Majorana vortex modes can emerge when the parent material in proximity to an $s$-wave superconductor is a 3D topological crystalline insulator with multiple surface Dirac cones protected by crystal symmetry. For instance, an even number of Dirac cones on the surface of 3D topological crystalline insulators (TCIs) such as SnTe~\cite{Hsieh12,Tanaka12experimental,SYXu12observation,Dziawa12} can theoretically host multiple Majorana fermions when $s$-wave superconductivity with a vortex is induced~\cite{CFang14,Liu14,Shiozaki14,Sato14mirror}.
More generally, $n$-fold rotation (C$_{n}$) symmetry-protected 3D TCIs with $n=2,4,6$ have
$n$ Dirac cones, instead of $2n$ Dirac cones, due to surface rotation anomaly on the top and bottom surfaces that are perpendicular to the rotation axis~\cite{ZSong17,ZSong18quantitative,Khalaf18,CFang19,Ahn20unconventional}. This motivates us to study interplay of the rotation anomaly and Majorana vortex zero modes on the surface of 3D TCIs with $s$-wave pairing, and, moreover, to establish general classification of crystal symmetry-protected Majorana vortex zero modes. 

In this paper we show that double Majorana fermions bound to a vortex are stable against symmetry-allowed perturbations in superconducting TCIs with C$_n$ rotation symmetry ($n=2,4,6$), and that their topological classification is extended from $\mathbb{Z}_2$ to $\mathbb{Z}_2 \times \mathbb{Z}_2$ accordingly. 
We consider a model of $s$-wave superconducting surface Dirac cones protected by C$_n$ symmetry, which is motivated from Ref.~\onlinecite{CFang19}, and show that double Majorana zero modes are bound to a vortex that preserves C$_n$ symmetry ($n=2,4,6$).
Using a lattice model of 3D high-spin TCIs with C$_2$ symmetry and $s$-wave Cooper pairing, we then show that the two surface Dirac cones from electrons with total angular momentum $J=1/2$ and $3/2$ accommodate double Majorana fermions at each end of a vortex line that is parallel to the rotation axis. With increasing chemical potential, the double Majorana fermions disappear successively at two vortex phase transitions that are distinguished by C$_2$ eigenvalues. 
Finally, we develop topological classification of Majorana zero modes bound to the ends of a vortex line under crystal symmetry. Our results include a new class of Majorana vortex zero modes classified by $\mathbb{Z} \times \mathbb{Z}$ under C$_{nv}$ symmetry (consisting of C$_n$ and a vertical-plane mirror-reflection), in addition to the $\mathbb{Z}_2\times\mathbb{Z}_2$ classification under C$_n$ symmetry.  Interestingly, under two-dimensional (2D) point groups, we find a one-to-one correspondence between topological classification of 3D TCIs and that of Majorana vortex zero modes, which implies that any symmetry-protected surface Dirac cone can accommodate a Majorana vortex zero mode when $s$-wave pairing is induced on the surface of TCIs.

\paragraph*{Surface rotation anomaly and Majorana vortex modes.--} 
To see the relation between Majorana vortex modes and C$_n$ symmetry-protected Dirac cones due to surface rotation anomaly, we first consider a 2D system of surface Dirac fermions on the surface of a 3D TCI with spin-orbit coupling, time-reversal symmetry (TRS), and C$_n$ symmetry for $n=2,4$, or 6.
A minimal model Hamiltonian for the surface of a 3D TCI with rotation anomaly is written as~\cite{CFang19}
\begin{equation}
\hat{H}^{\rm surf} = \sum_{|\bm{k}| < \Lambda} \sum_{s,s',\sigma, \sigma'} \hat{c}_{\bm{k},s,\sigma}^{\dagger} H_{s,\sigma; s',\sigma'}(\bm{k})\hat{c}_{\bm{k},s',\sigma'}
\end{equation}
with
\begin{align}
 H(k_x,k_y) = v (k_x s_x +  k_y s_y ) \otimes \bm{1}_2, \label{eq:surfaceDirac}
\end{align}
where $s_i$ ($i=x,y,z$) are the Pauli matrices in the spin space, $s$ and $s'$ are spin indices ($s \in \{\uparrow,\downarrow\}$), $\sigma$ and $\sigma'$ are orbital indices ($\sigma \in \{1,2\}$), $\bm{1}_n$ the $n \times n$ identity matrix, $v$ is the velocity ($v>0$), and $\Lambda$ is a cutoff of the order of the inverse lattice spacing.
In this minimal model the rotation axis is along the $z$ axis and the double Dirac cones are centered at $\bar\Gamma$ point $(k_x,k_y)=(0,0)$. The Dirac point can be split into $n$ Dirac points away from the $\bar{\Gamma}$ point by adding symmetry-allowed perturbations~\cite{CFang19}.  Equation~(\ref{eq:surfaceDirac}) satisfies time-reversal symmetry (TRS), $T H(\bm{k}) T^{\dagger} = H(-\bm{k})$ with $T = is_y K$, and C$_n$ symmetry,
$C_n  H(\bm{k}) C_n^{\dagger} = H(R_n \bm{k})$
with
\begin{align}
 C_n = e^{-i \frac{ \pi}{n} s_z} \otimes \sigma_z, 
\label{eq:rotation}
\end{align}
where $K$ is complex conjugation, $\sigma_i$ ($i=x,y,z$) are the Pauli matrices in the orbital space, and
$R_n$ is a representation of $O(2)$, e.g., $R_4: (k_x,k_y) \to (-k_y,k_x) $.
The orbitals are assumed to have opposite parities (e.g., $s$ and $p$ orbitals), as indicated by $\sigma_z$ in Eq.\ (\ref{eq:rotation}). 
The double Dirac cones belong to different irreducible representations of C$_n$ and cannot be continuously deformed into each other while preserving C$_n$ symmetry.
Thus the Dirac cones cannot be gapped out; for example, a TRS-preserving mass term $s_z \otimes \sigma_y$ is prohibited since $\{C_n,s_z\otimes\sigma_y\}=0$.

Suppose that an $s$-wave superconductor is deposited on the rotation-invariant surface of the TCI so that the $s$-wave Cooper pairs are induced due to the proximity effect. The surface Hamiltonian $\hat{H}^\mathrm{surf}$ is extended to the Bogoliubov-de Gennes (BdG) Hamiltonian
\begin{align}
\hat{H}^{\rm surf}_{\rm BdG} =&
\sum_{|\bm{k}| < \Lambda} \Big\{ \sum_{s,s',\sigma, \sigma'} \hat{c}_{\bm{k},s,\sigma}^{\dagger}
 [H(\bm{k})-\mu \bm{1}_4 ]_{s,\sigma;s',\sigma'}
\hat{c}_{\bm{k},s',\sigma'} \notag \\
&\quad
+  \left[\Delta_0 \!
\left(\hat{c}_{\bm{k},\uparrow, 1}^{\dagger} \hat{c}_{-\bm{k},\downarrow, 1}^{\dagger}
+ \hat{c}_{\bm{k},\uparrow, 2}^{\dagger}\hat{c}_{-\bm{k},\downarrow, 2}^{\dagger}\right)
+ {\rm H.c.}\right]\Big\}, \label{eq:surfacBdG}
\end{align}
where $\mu$ is the chemical potential and $\Delta_0$ is the induced s-wave superconducting gap, which is assumed to have the same magnitude for the two orbitals for simplicity.
The BdG Hamiltonian $\hat{H}^\mathrm{surf}_\mathrm{BdG}$ describes a C$_n$-symmetric fully-gapped superconductor with the energy spectrum 
$E_{\bm{k}} = \pm\sqrt{(\pm v |\bm{k}|-\mu)^2 + |\Delta_0|^2}$ when $\Delta_0 \neq 0$.

In the presence of a superconducting vortex at the rotation axis $(x,y)=(0,0)$,
the order parameter $\Delta$ takes the form
$\Delta (r) e^{i \theta}$, where $r=\sqrt{x^2+y^2}$ and $\theta = \arctan(y/x)$.
Here $\Delta(r)$ is a monotonic function of $r$ satisfying $\Delta(0)=0$ and $\Delta(\infty)=\Delta_0$.
We note that the vortex is placed on the rotation axis, and this assumption is naturally satisfied when the superconducting coherence length is much larger than the size of a unit cell.
The existence of a vortex breaks TRS, whereas C$_n$ symmetry still holds, albeit in a modified form, since a vortex field also rotates as $ \Delta e^{i \theta} \to \Delta e^{i \left(\theta + \frac{2 \pi}{n}\right)}$ under C$_n$. Thus, the correct form of C$_n$ operations in the presence of a vortex~\cite{Qin19} is Eq.~(\ref{eq:rotation}) combined with a gauge transformation by $e^{i \pi/n}$,
\begin{align}
 c_{\bm{k},s,\sigma}^{\dagger} \to  \sum_{s' \sigma'}c_{R_n \bm{k}, s',\sigma'}^{\dagger} [C_n]_{s' \sigma'; s \sigma} \; e^{i \frac{\pi}{n}}. \label{eq:rotationBdG}
 \end{align}
One can easily verify that $\hat{H}^\mathrm{surf}_\mathrm{BdG}$ is invariant under the transformation of Eq.~(\ref{eq:rotationBdG}).
The gauge transformation changes the eigenvalues of the C$_n$ operation from the double (spinful) values to the single (spinless) values, since $e^{i \frac{\pi (2m-1)}{n} + i \frac{\pi}{n} } = e^{i \frac{2\pi m}{n} }$ ($m = 1,2, \cdots, n$). Note that particle-hole symmetry (PHS) is respected by the BdG Hamiltonian even in the presence of a vortex. 

The BdG Hamiltonian with a vortex is equivalent to a 2D Dirac Hamiltonian coupled with the $s$-wave pairing, which is known as the Jackiw-Rossi model~\cite{JackiwRebbi1976,JackiwRossi1981,FuKane08,Fukui10}. 
In particular, replacing $\bm{k}$ with $-i \bm{\partial}$, we can solve the BdG Hamiltonian analytically and obtain two zero-energy solutions, which take particularly simple forms at $\mu=0$:
\begin{subequations}
\label{eq:doubleMFs}
\begin{align}
 &\hat{\gamma}_1 = \int d \bm{x} \; \left(e^{i \frac{\pi}{4}}\hat{c}_{\uparrow,1}(\bm{x}) + e^{-i \frac{\pi}{4}}\hat{c}_{\uparrow,1}^{\dagger}(\bm{x})\right) e^{- \int^r_0 \Delta (r')/v d r'}, \\
 &\hat{\gamma}_2 = \int d \bm{x} \; \left(e^{i \frac{\pi}{4}} \hat{c}_{\uparrow,2}(\bm{x}) +e^{-i \frac{\pi}{4}} \hat{c}_{\uparrow,2}^{\dagger}(\bm{x})\right) e^{- \int^r_0 \Delta (r')/v d r'},
 \end{align}
\end{subequations}
where $\hat{c}_{s,\sigma}(\bm{x}) $ is the Fourier transformation of $\hat{c}_{\bm{k},s,\sigma}$. Equations~(\ref{eq:doubleMFs}) satisfy the Majorana condition $\gamma_{1(2)}^{\dagger} = \gamma_{1(2)}$. We note that reversing the sign of the vortex phase leads to zero-energy states with spin-down electrons, and the zero-energy solutions for $\mu \neq 0$ involve both spin components~\cite{suppl}.  

To check the stability of the Majorana vortex zero modes, we examine possible perturbations to Majorana fermions.
In general the coupling between Majorana fermions can be written as
\begin{align}
\hat{H}_{\rm MF} = \sum_{a,b} i \mathcal{A}_{ab} \hat{\gamma}_a \hat{\gamma}_b, \label{eq:HMF}
\end{align}
where $\mathcal{A}_{ab}$ is a real-skew matrix and $\hat{\gamma}_a^{\dagger} = \hat{\gamma}_a$.
If such coupling is allowed by symmetry, Majorana zere modes acquire a finite hybridization gap.
For the two Majorana zero modes in Eq.\ (\ref{eq:doubleMFs}),
however, the coupling is prohibited by C$_n$ symmetry, since Majorana zero modes are eigenstates, transformed as $(\hat{\gamma}_1,\hat{\gamma}_2) \to (\hat{\gamma}_1,-\hat{\gamma}_2)$, of the C$_n$ operation defined in Eq.\ (\ref{eq:rotationBdG});
$\hat{\gamma}_1$ and $\hat{\gamma}_2$ belong to the different sector of C$_n$ symmetry. 
Furthermore, the topological classification is found to be $\mathbb{Z}_2 \times \mathbb{Z}_2$ as follows. When we extend the BdG Hamiltonian to double ones $\hat{H}_\mathrm{BdG}^\mathrm{surf} \oplus \hat{H}_\mathrm{BdG}^\mathrm{surf}$, we can find four Majorana fermions $\hat{\gamma}_a$ $(a=1,\ldots,4)$ that obey the transformation $(\hat{\gamma}_1, \hat{\gamma}_2, \hat{\gamma}_3, \hat{\gamma}_4) \to (\hat{\gamma}_1, -\hat{\gamma}_2, \hat{\gamma}_3, -\hat{\gamma}_4)$ under the C$_n$ operation. In this case, a symmetry-preserving hybridization term $\hat{H}_{\rm MF} =i \lambda \left( \gamma_1 \gamma_3 + \gamma_2 \gamma_4 \right)$ is allowed, since two out of four Majorana fermions share the same eigenvalues of C$_n$ symmetry.

 \begin{figure}[tbp]
\centering
 \includegraphics[width=7cm]{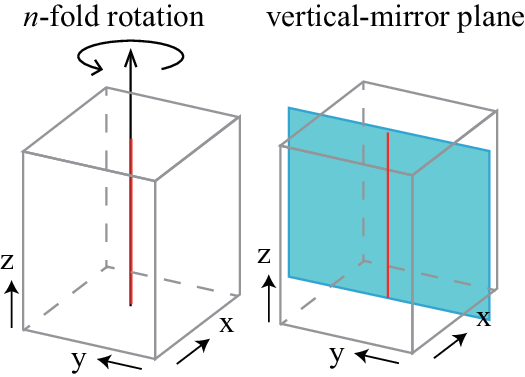}
 \caption{Schematic illustration of $n$-fold rotation and vertical-mirror-reflection operations in a 3D SC with a vortex line (red line).}\label{fig:symmetry}
\end{figure}

\paragraph*{Symmetry-protected vortex phase transition in a lattice model.--} 
So far we have focused on the surface effective Hamiltonian.  This approach is valid for 3D superconducting TCIs where the 3D bulk is insulating.  Now we consider a 3D C$_2$-invariant lattice model with a vortex line in order to show that double Majorana vortex zero modes can appear at each end of the vortex line, even when the bulk is doped into a metallic state.
As the chemical potential increases, a pair of zero modes localized at the opposite ends are expected to have a longer localization length along the vortex line and pair-annihilate at a critical point $\mu=\mu_{\rm c}$, which is known as a vortex phase transition~\cite{Hosur11}. Here we show the existence of two successive vortex phase transitions~\footnote{Incidentally, a different type of multiple vortex phase transitions has been proposed in higher-order TIs~\cite{Ghorashi20}. } that are protected by C$_2$ symmetry.

To this end, we introduce a model Hamiltonian of high-spin fermions in the normal state, 
\begin{align}
H(\bm{k}) =& \left( M + m_0 \sum_{i=x,y,z} \cos (k_i)\right) \bm{1}_4 \otimes \tau_z \notag \\
 & + t \sum_{i=x,y,z} \sin (k_i) J_i \otimes \tau_x \notag \\
 & +\left[\delta_x \sin (k_x) J_y + \delta_y \sin (k_y) J_x \right] \otimes \tau_x \notag \\
 & +\delta' (J_xJ_y+J_yJ_x) \otimes \tau_x , \label{eq:doubleTI}
\end{align}
where $J_i$ are the $4 \times4$ spin matrices in the spin-$3/2$ representation, and $\tau_i$ the Pauli matrices in the orbital space.
In addition to the parameters $M$, $m_0$, and $t$, we have introduced lattice distortions $\delta_x$ and $\delta_y$ to break C$_4$ symmetry down to C$_2$ symmetry ($C_2 = e^{i \pi J_z} \otimes \tau_0$) and distortion $\delta'$ to break C$_4$ and inversion ($\bm{1}_4\otimes \tau_z$) symmetries.  The model~(\ref{eq:doubleTI}) may be realized in the antiperovskite compounds~\cite{Kariyado11letter,Kariyado12full,Hsieh14,Oudah16,Kawakami18}, in which two $\Gamma_8$ bands with different orbital characters are formed,  by virtue of spin-orbit coupling and cubic symmetry, around the $\Gamma$ point, where the band inversion leads to a TCI phase. When $\delta'=0$, the Hamiltonian $H(\bm{k})$ in Eq.~(\ref{eq:doubleTI}) has the double band inversions at the $\Gamma$ point in the parameter regime $-3 < M/m_0 < -1$, and the 2D surface has two gapless modes protected by TRS ($T= e^{i \pi J_y}\otimes \tau_0K$) and C$_2$ symmetry~\cite{suppl}, where $C_2$ projected onto the surface states has a similar form to Eq.~(\ref{eq:rotation}). 
We implement $s$-wave pairing to Eq.~(\ref{eq:doubleTI}), with the gap function $\Delta(\bm{x}) = \Delta_0 \tanh \left( r/\xi \right) e^{i \theta}$ in the cylindrical coordinates with the coherence length $\xi$ and the vortex line on the rotation axis; see Fig.~\ref{fig:symmetry} (left).

We numerically diagonalize the 3D BdG Hamiltonian with the vortex line under the periodic boundary condition (PBC) in the $z$ direction and open boundary conditions in the $x$ and $y$ directions, and obtain the energy spectrum of quasiparticles.  Figure~\ref{fig:doubleTI} (a) shows the energy spectrum at momentum $k_z=0$ as a function of the chemical potential $\mu$.  The energy levels within the energy gap $\Delta_0$ are (Caroli-de Gennes-Matricon\cite{Caroli1964}) bound states in the vortex. As expected, level crossings at $E=0$ occur twice, at $\mu_{\rm c, 1} \simeq 0.62$ and $\mu_{\rm c, 2} \simeq 0.87$, in Fig.~\ref{fig:doubleTI} (a), signaling two vortex phase transitions at which a pair of Majorana zero modes from opposite surfaces annihilate; see also Figs.~\ref{fig:doubleTI} (b) and~S1 (b) in Ref.~\onlinecite{suppl}.
 We note that the two vortex phase transitions are distinguishable in terms of the C$_2$ eigenvalues, meaning that the Majorana vortex end modes and associated vortex phase transitions are protected by the C$_2$ symmetry.

\begin{figure}[tbp]
\centering
 \includegraphics[width=7.5cm]{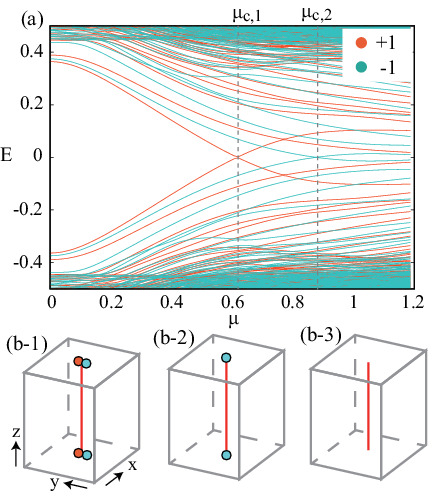}

 \caption{
(a) Evolution of energy levels of vortex bound states as a function of the chemical potential at the $k_z=0$ plane in the $s$-wave superconducting state of (\ref{eq:doubleTI}) under PBC in the $z$ direction. We assume $(M,m_0,t, \delta_x,\delta_y,\delta') = (2.5,-1,1.5,0.1,0.2,0.1)$, $\Delta_0 = 0.5$, $\xi=4$, and the lattice sizes $L_x=L_y=21$.
The energy levels in red (blue) have $C_2$ eigenvalues $+1$ ($-1$). 
(b) If open boundary condition is imposed (instead of PBC) in the $z$ direction, the number of Majorana states localized at one end of a vortex line is (b-1) two for $0\le\mu<\mu_{\rm c,1}$, (b-2) one for $\mu_{\rm c,1}<\mu<\mu_{\rm c,2}$, and (b-3) none for $\mu>\mu_{\rm c, 2}$.
}\label{fig:doubleTI}
\end{figure}

\paragraph*{Topological classification of Majorana vortex end modes.--} 
In the remaining part of this paper, we discuss topological classification of Majorana vortex end modes in superconducting 3D TCIs. The vortex phase transitions are related to the change in the topology of electronic states in a vortex line. The relevant energy scale of these states is the level spacing of vortex bound states, i.e., a mini gap~\cite{Caroli1964} $\delta \sim \Delta/(k_{\rm F} \xi) \ll \Delta$ ($k_{\rm F}$ is the Fermi wave number).  The bulk excitations with energy $E\gtrsim\Delta$ are irrelevant in our discussion, and we are allowed to take a finite system size in the directions perpendicular to the vortex line.
Since a vortex line breaks TRS, our problem is reduced to classification of quasi-one-dimensional (Q1D) SCs in class D of the Altland-Zirnbauer (AZ) classes~\cite{Altland97,Zirnbauer96,Schnyder08}. Thus, without crystalline symmetry, Q1D superconducting vortices are classified by $\mathbb{Z}_2$.

Crystal symmetries that can be preserved under the presence of a vortex line and a surface termination are 2D point groups C$_{n}$ or C$_{nv}$ ($n=1,2,3,4,6$), when a vortex line is on the rotation axis or the vertical-mirror plane (see Fig.~\ref{fig:symmetry}).
The topological classification of Q1D SCs with point group symmetry has been discussed in the previous works~\cite{CFang17,Cornfeld19,Shiozaki19classification}.
We here employ those approaches and show that the topological classification is modified by the presence of the $U(1)$ vortex field. To this end, we start from BdG Hamiltonian describing Q1D $s$-wave SCs with an infinitely long vortex line along the $z$ axis, $\mathcal{H} (k_z)$, in which the finite lattice sites in the $xy$ plane are implicitly included as sublattice degrees of freedom. 
The BdG Hamiltonian has PHS, $\mathcal{P} \mathcal{H} (k_z) \mathcal{P}^{-1} = -\mathcal{H} (-k_z)$ with $\mathcal{P}^2 =\bm{1}_{N}$, where $N$ is the dimension of $\mathcal{H}(k_z)$. 
The C$_n$ symmetry ($n=2,3,4,6$) imposes the constraint $\mathcal{C}_n \mathcal{H} (k_z) \mathcal{C}_n^{-1} = \widetilde{\mathcal{H}} (k_z)$, where $\widetilde{H}(k_z)$ is related to $\mathcal{H}(k_z)$ by C$_n$ rotation of lattice sites in the $xy$ plane.
Furthermore, we can set $[\mathcal{C}_n,\mathcal{P}]=0$ with $s$-wave Cooper pairing.  The gauge transformation associated with the U(1) vortex field as discussed above Eq.~(\ref{eq:rotationBdG}) leads to $(\mathcal{C}_n)^n =\bm{1}_{N}$. As a result, the BdG Hamiltonian can be block-diagonalized as
\begin{align}
 \mathcal{H}_{0} \oplus \mathcal{H}_{1}  \oplus \cdots \oplus \mathcal{H}_{n-1},   \label{eq:subsectors}
\end{align}
where $\mathcal{H}_m$ is a Hamiltonian in the subsector with $\mathcal{C}_n$-eigenvalue $e^{i 2 \pi m/n}$. Since C$_n$ symmetry forbids any mixing of states from different subsectors, we can define an AZ symmetry class for each subsector. 
When $m/n \notin \{ 0, 1/2 \}$, the eigenvalues are complex numbers. In this case $\mathcal{H}_m$ does not have PHS itself and belongs to class A. On the other hand, when $m/n \in \{ 0, 1/2 \}$, the $\mathcal{C}_n$-eigenvalues are real numbers, and PHS remains as a symmetry of $\mathcal{H}_m$, meaning that the subsector is in class D. According to the periodic table of topological insulators/superconductors~\cite{Schnyder08,Schnyder09,Kitaev09,Ryu10}, Q1D SCs are classified by $\mathbb{Z}_2$ for class D and $0$ for class A. Thus, the topological classification under C$_n$ symmetry becomes  $\mathbb{Z}_2 \times \mathbb{Z}_2$ for $n=2,4,6$ and $\mathbb{Z}_2$ for $n=3$, which are consistent with the results from the effective surface theory discussed above. 
The topological indices $(\nu_{+},\nu_{-}) \in \mathbb{Z}_2 \times \mathbb{Z}_2$ are given by
\begin{align}
\nu_{\pm} = 
\sgn\!\left\{ \pf [ \mathcal{U}_{\pm} \mathcal{H}_{\pm}(k_z=0) ]
                   \pf [ \mathcal{U}_{\pm} \mathcal{H}_{\pm}(k_z=\pi) ]\right\},
\label{eq:z2inv}
\end{align}
where $\mathcal{H}_{+}$ ($\mathcal{H}_{-}$) is Hamiltonian in the subsectors with $\mathcal{C}_n$-eigenvalues $+1$ ($-1$), and $\mathcal{U}_{\pm}$ is the unitary part of $\mathcal{P}$ projected to $\mathcal{H}_{\pm}$.

\begin{table}[tb]
\caption{
Classification of Majorana vortex end modes under 2D point groups. The first, second, and third columns represent 2D point groups, relevant AZ symmetry classes, and 1D topological invariants, respectively. Here, D$^2$ stands for D $\times$ D and so too with the others.
}
\label{tab:classification}
\begin{tabular}{ccc}
\hline\hline
Symmetry & AZ class & $1$ dim.\\
\hline 
C$_{1}$          & D &  $\mathbb{Z}_2$ \\ 
C$_{2}$ & D$^2$ & $\mathbb{Z}_2 \times \mathbb{Z}_2$\\
C$_{3}$ &  D $\times$ A$^2$ & $\mathbb{Z}_2 $\\
C$_{4}$ & D$^2$ $\times$ A$^2$ & $\mathbb{Z}_2 \times \mathbb{Z}_2$\\
C$_{6}$ & D$^2$ $\times$ A$^4$ & $\mathbb{Z}_2 \times \mathbb{Z}_2 $\\
C$_{1v}$       & BDI  & $\mathbb{Z} $\\
C$_{2v}$ & BDI$^2$ & $\mathbb{Z} \times \mathbb{Z}$\\
C$_{3v}$ & BDI $\times$ AI$^2$ & $\mathbb{Z} $\\
C$_{4v}$ & BDI$^2$ $\times$ AI$^2$ & $\mathbb{Z} \times \mathbb{Z} $\\
C$_{6v}$ & BDI$^2$ $\times$ AI$^4$ & $\mathbb{Z} \times \mathbb{Z} $\\
\hline\hline
\end{tabular} 
\end{table}

Next, we consider the effect of vertical-mirror-reflection (C$_{1v}$) symmetry. Following the arguments in Refs.~\onlinecite{CFang14,Shiozaki14}, we find that the BdG Hamiltonian $\mathcal{H}(k_z)$ with a vortex is invariant not by C$_{1v}$ transformation but by magnetic-mirror transformation $\mathcal{M}_{\rm T}=T C_{1v}$, which is the combination of C$_{1v}$ and time-reversal transformation, as $\mathcal{M}_{\rm T} \mathcal{H}(k_z)\mathcal{M}_{\rm T}^{-1}=\mathcal{H}(-k_z)$.
For $s$-wave pairing we can take $[\mathcal{M}_{\rm T},\mathcal{P}]=0$. Here, $\mathcal{M}_{\rm T}$ is antiunitary and $\mathcal{M}_{\rm T}^2 = \bm{1}_N$. 
Thus, $\mathcal{M}_{\rm T}$ plays a role of TRS for spinless fermions in the BdG Hamiltonian, and $\mathcal{H}$ belongs to class BDI with the 1D topological invariant $\mathbb{Z}$.

Finally, we consider C$_{nv}$ symmetry ($n=2,3,4,6$) that consists of C$_n$ and vertical-mirror reflections, where the rotation axis is in the vertical-mirror planes (see Fig.~\ref{fig:symmetry}). The two operations do not commute, and the rotation direction of C$_n$ is inverted by the vertical-mirror reflection.  The time-reversal operation $T$ also affects $\mathcal{C}_n$ by changing the phase factor in the gauge transformation in Eq.~(\ref{eq:rotationBdG}) to its complex conjugate. As a result, $\mathcal{C}_n$ and $\mathcal{M}_{\rm T}$ satisfy the relation
\begin{align}
 \mathcal{M}_{\rm T}\mathcal{C}_n\mathcal{M}_{\rm T}^{-1} = \mathcal{C}_n^{-1}. \label{eq:magneticmirror}
\end{align}
From Eq.~(\ref{eq:magneticmirror}) and the anti-unitarity of $\mathcal{M}_{\rm T}$, we readily find that $\mathcal{M}_{\rm T}$ is closed within each subsector in Eq.~(\ref{eq:subsectors}). Thus, $\mathcal{M}_{\rm T}$ can be regarded effectively as (spinless) TRS in each $\mathcal{H}_m$. Hence, $\mathcal{H}_m$ belongs to class AI when $m/n \notin \{ 0, 1/2 \}$, while it is in class BDI when $m/n \in \{ 0, 1/2 \}$. Therefore, 1D topological invariants for $\mathcal{H}$ are $\mathbb{Z} \times \mathbb{Z}$ for $n=2,4,6$ and $\mathbb{Z}$ for $n=3$. Since the combination of the spinless TRS and PHS gives a chiral symmetry, we can define winding numbers $(w_+,w_-) \in \mathbb{Z} \times \mathbb{Z}$ as
\begin{equation}
 w_{\pm} = \frac{1}{4\pi i} \int d k_z \rm Tr \left[ \Gamma_{\rm M} \mathcal{H}_{\pm}(k_z)^{-1} \partial_{k_z} \mathcal{H}_{\pm}(k_z) \right],
\end{equation}
where $\Gamma_{\rm M}$ is a chiral operator, defined by $M_{\rm T} \mathcal{P}$ projected onto $\mathcal{H}_{\pm}$. The classification of Majorana vortex end modes is summarized in Table~\ref{tab:classification}. Interestingly, our classification has one-to-one correspondence with that of 3D TCIs with 2D point groups~\cite{Cornfeld19}.
The correspondence can be understood from our analysis of 2D surface theory~(\ref{eq:surfacBdG}), as crystal symmetry-protected multiple surface Dirac cones are able to host multiple Majorana vortex zero modes via the Fu-Kane mechanism~\cite{FuKane08}. These multiple Majorana vortex zero modes are eigenstates of 2D point groups and free from a hybridization. 

\paragraph*{Concluding remarks.--} 
We have studied the intrinsic relation between surface rotational anomaly and Majorana zero modes localized at the ends of a vortex in 3D superconductors, and established topological classification predicting the existence of double Majorana vortex zero modes: $\mathbb{Z}_2 \times \mathbb{Z}_2$ for C$_2$, C$_4$, and C$_6$, and $\mathbb{Z} \times \mathbb{Z}$ for C$_{2v}$, C$_{4v}$, and C$_{6v}$. 
The double Majorana zero modes can be realized, e.g., in high-spin topological insulators~\cite{Oudah16}, SnTe~\cite{Sasaki12,TSato13}, BiBr~\cite{Tang2019}, and a family of Zintl compounds~\cite{Zhang19}, and detected through tunneling conductance as a zero bias conductance peak of height $4e^2/h$. 

\begin{acknowledgments}
S.K. thanks Masatoshi Sato and Yuki Kawaguchi for valuable discussions. 
This work was supported by JSPS KAKENHI (Grant Nos.\ 19K03680,  19K14612) and JST CREST (Grant Nos.\ JPMJCR16F2, JPMJCR19T2).
\end{acknowledgments}

\bibliography{Majorana_vortex}

\newpage

\widetext

\renewcommand{\thefigure}{S\arabic{figure}} 

\renewcommand{\thetable}{S\arabic{table}} 

\renewcommand{\thesection}{S\arabic{section}.}

\renewcommand{\theequation}{S.\arabic{equation}}

\setcounter{figure}{0}
\setcounter{table}{0}
\setcounter{equation}{0}

\begin{center} 
{\large {\bf Supplementary materials: \\ Double Majorana vortex zero modes in superconducting topological crystalline insulators with surface rotation anomaly}}
\end{center}



\begin{flushleft} 
{\bf S1. Majorana vortex zero modes at $\mu \neq 0$}
\end{flushleft}

In the main text, we have presented the explicit form of Majorana vortex zero modes at $\mu=0$ only. We discuss here the zero-energy solution of the BdG Hamiltonian at $\mu \neq 0$, and show that those Majorana vortex zero modes remain stable under $n$-fold rotation symmetry. 

To begin with, we consider a model of a single surface Dirac cone in proximity to an s-wave superconductor with a vortex  $\Delta \to \Delta(r) e^{i \theta}$,
\begin{align}
 H(\bm{x}) = 
 \begin{pmatrix}
  -\mu & -i v(\partial_x -i \partial_y) & 0 & \Delta(r)e^{i \theta} \\ 
   -i v(\partial_x +i \partial_y) & -\mu &  -\Delta(r)e^{i \theta}& 0\\
   0 & -\Delta(r)e^{-i \theta} & \mu & i v(\partial_x +i \partial_y) \\
   \Delta(r)e^{-i \theta} & 0 & i v(\partial_x -i \partial_y) & \mu
 \end{pmatrix}, 
\end{align}
where we suppose $\hbar =1$, $v >0$, $\Delta(r)$ is a monotonic function that satisfies $\Delta(0)=0$ and $\Delta(\infty)=\Delta_0$, $r=\sqrt{x^2+y^2}$, and $\theta = \arctan (y/x)$. The problem boils down to the eigenvalue problem: $H(\bm{x}) \phi(\bm{x})= E \phi(\bm{x})$ with $\phi(\bm{x}) = (u_{\uparrow}(\bm{x}), u_{\downarrow}(\bm{x}), v_{\uparrow}(\bm{x}), v_{\downarrow}(\bm{x}))^T$. In particular, when $E=0$, $v_{\uparrow}(\bm{x}) = u_{\uparrow}^{\ast}(\bm{x})$ and $v_{\downarrow}(\bm{x}) = u_{\downarrow}^{\ast}(\bm{x})$ are satisfied due to particle-hole symmetry. Thus, we have a couple of equations:
\begin{align}
& -\mu u_{\uparrow}(r, \theta) + e^{-i \theta}v(-i\partial_r - \partial_{\theta}/r) u_{\downarrow}(r, \theta) + e^{i \theta}\Delta(r) u_{\downarrow}^{\ast}(r, \theta) =0, \notag \\
& -\mu u_{\downarrow}(r, \theta) + e^{i \theta}v(-i\partial_r + \partial_{\theta}/r) u_{\uparrow}(r, \theta) - e^{i \theta}\Delta(r) u_{\uparrow}^{\ast}(r, \theta) =0 , \label{eq:eigen}
\end{align} 
where we use the the polar coordinate, $\partial_x \pm i\partial_y =e^{\pm i \theta} (\partial_r \pm i \partial_{\theta}/r)$. To find the solution, we assume the form of $u_{\uparrow}(r,\theta)$ and $u_{\downarrow}(r,\theta)$ as~\cite{Chamon10}
\begin{align}
& u_{\uparrow}(r,\theta) = f(r) \exp \left[-i \frac{\pi}{4} - \int^r_0 \Delta(r')/v \, dr' \right], \notag \\
& u_{\downarrow}(r,\theta) = g(r) \exp \left[i \left( \theta + \frac{\pi}{4} \right)- \int^r_0 \Delta(r')/v \, dr' \right], \label{eq:ansatz}
\end{align}
where $f(r)$ and $g(r)$ are real functions of $r$. Substituting Eq.~(\ref{eq:ansatz}) into Eq.~(\ref{eq:eigen}) yields 
\begin{align}
 &\partial_r (r g(r)) = r (\mu/v) f(r), \notag \\
 &\partial_r f(r) = - (\mu/v) g(r).
\end{align}
Thus, we find that $f(r) \propto J_0((\mu/v) \, r)$ and  $g(r) \propto J_1((\mu/v) \, r)$, where $J_\nu (x)$ is the Bessel function of the first kind. As a result, the zero-energy solution with nonzero $\mu$ is of the form:
\begin{align}
 \phi_0(r , \theta) = N \, \left(\begin{array}{@{\,} c @{\,}} J_0((\mu/v) \, r) e^{-i \pi/4} \\ J_1((\mu/v) \, r) e^{i (\theta + \pi/4)} \\ J_0((\mu/v) \, r) e^{i \pi/4} \\ J_1((\mu/v) \, r) e^{-i (\theta + \pi/4)} \end{array} \right) e^{- \int^r_0 \Delta(r')/v \, dr'}, \label{eq:sol_Dirac}
\end{align}
where $N$ is a normalization constant. We note that when $\mu \neq 0$, the spin up and down components coexist.

We now turn to the case of double surface Dirac cones, whose BdG Hamiltonian is described as $H'(\bm{k})=H(\bm{k}) \oplus H(\bm{k})$. From (\ref{eq:sol_Dirac}) the zero-energy solutions are readily obtained as 
\begin{align}
&\phi_1(r , \theta) = \phi_0(r , \theta) \oplus \bm{0}, \notag \\
&\phi_2(r , \theta) = \bm{0} \oplus \phi_0(r , \theta),  
  \label{eq:sol_Dirac_double}
\end{align}
where $\bm{0}$ is the null vector. 
Using the zero-energy solutions, the Marajoana operators are defined by 
\begin{align}
\gamma_i &\equiv \int d \bm{x} \; \phi^{\dagger}_{i}(\bm{x}) \hat{\Psi} (\bm{x}) \notag \\
        &=\sum_{s, \sigma}\int d \bm{x} \left\{ u_{s, \sigma, i}^{\ast} (\bm{x}) c_{s, \sigma}(\bm{x})+ v_{s, \sigma, i}^{\ast} (\bm{x}) c^{\dagger}_{s,\sigma}(\bm{x}) \right\}, \ \ (i=1,2),
\end{align} 
where $\hat{\Psi}_{s, \sigma}(\bm{x}) = (\hat{c}_{s, \sigma}(\bm{x}), \hat{c}_{s, \sigma}^{\dagger}(\bm{x}))^T$ and the indices $s$ and $\sigma$ represent the spin and orbital spaces, respectively. 

We define the C$_n$ transformation ($n=2,4,6$) such that $\hat{c}_{s,\sigma}^{\dagger} (\bm{x}) \to \sum_{s',\sigma'} \hat{c}_{s' \sigma'}^{\dagger}(R_n \bm{x}) [C_n]_{s' \sigma' ;s\sigma} \; e^{i \pi/n}$ (see Eq.~(5) in the main text). Thus, the C$_n$  operation in the Nambu space is described as
\begin{align}
 \mathcal{C}_n = \begin{pmatrix} e^{i \frac{\pi}{n} }C_n & 0 \\ 0 & e^{-i \frac{\pi}{n} }C_n^{\ast} \end{pmatrix}, \ \ C_n = e^{- i \frac{\pi}{n} s_z} \otimes \sigma_z, \label{eq:rot_op}
\end{align} 
and $\hat{\Psi}_{s,\sigma}^{\dagger} (\bm{x}) \to \sum_{s',\sigma'} \hat{\Psi}_{s',\sigma'}^{\dagger}(R_n \bm{x}) [\mathcal{C}_n]_{s'\sigma' ;s \sigma } $ accordingly.
From Eqs.~(\ref{eq:sol_Dirac_double}) and (\ref{eq:rot_op}), we obtain
\begin{align}
 \mathcal{C}_n \phi_1(r, \theta) = \phi_1(r, \theta+2\pi/n) , \ \ \mathcal{C}_n \phi_2(r, \theta) = -\phi_2(r, \theta+2\pi/n). \label{eq: rot_wavefn}
\end{align}
Therefore, the Majorana operators are transformed, under  the C$_n$ transformation, as
\begin{align}
 \gamma_i &\to \int d \bm{x} \; \phi^{\dagger}_i(\bm{x})  \left( \mathcal{C}_n^{\dagger} \hat{\Psi}(R_n\bm{x}) \right) \notag \\
  &=  (-1)^{i+1} \int d \bm{x} \; \phi^{\dagger}_i(R_n \bm{x})  \hat{\Psi}(R_n\bm{x}) \notag \\
  &=  (-1)^{i+1} \gamma_i,
\end{align} 
where we have used Eq.~(\ref{eq: rot_wavefn}) in the second line.
The result leads to the stability of double Majorana vortex zero modes under symmetry-preserving perturbations as discussed in the main text.

\begin{flushleft} 
{\bf S2. Surface states and Majorana vortex zero modes in a model of high-spin fermions}
\end{flushleft}

We consider a model of high-spin fermions:
\begin{align}
H(\bm{k}) = \left(M + m_0 \sum_{i=x,y,z} \cos (k_i) \right) \bm{1}_4 \otimes \tau_z + t \sum_{i=x,y,z} \sin (k_i) J_i \otimes \tau_x + [\delta_x \sin(k_x) J_y + \delta_y \sin(k_y) J_x ] \otimes \tau_x, \label{eq:doubleTI}
\end{align}
where $J_i$ are the $4 \times4$ spin matrices in the spin-$3/2$ basis and $\tau_i$ the Pauli matrices in the orbital space. $M$, $m_0$, $t$, $\delta_x$, and $\delta_y$ are material parameters.  
As discussed in the main manuscript, the double band inversions occur at the $\Gamma$ point in the parameter regime $-3 < M/m_0 < -1$, leading to two Dirac cones on a surface. 
Introducing s-wave pairing with a vortex at $(x,y)=0$ into Eq. (\ref{eq:doubleTI}) yields two Majorana zero modes at each end of the vortex.
We here illustrate those topological states by numerically diagonalizing the normal-state and Bogoliubov-de Gennes Hamiltonians. In Fig.~\ref{fig:doubleTI} (a), we show the ($001$) surface states of Eq.~(\ref{eq:doubleTI}), which realize two Dirac cones (one of them has nonlinear dispersion). In Fig.~\ref{fig:doubleTI} (b), we plot eigenvalues of the BdG Hamiltonian for Eq.~(\ref{eq:doubleTI}) with an $s$-wave pairing hosting a vortex line along the $z$ axis. The gap function is given by $ \Delta_0 \tanh \left( r/\xi \right) e^{i \theta}$ in the polar coordinate, where $\xi$ is the coherence length. We find four zero energy states, which describe two Majorana vortex zero modes at each end of the vortex line.

\begin{figure}[bp]
\centering
 \includegraphics[width=10cm]{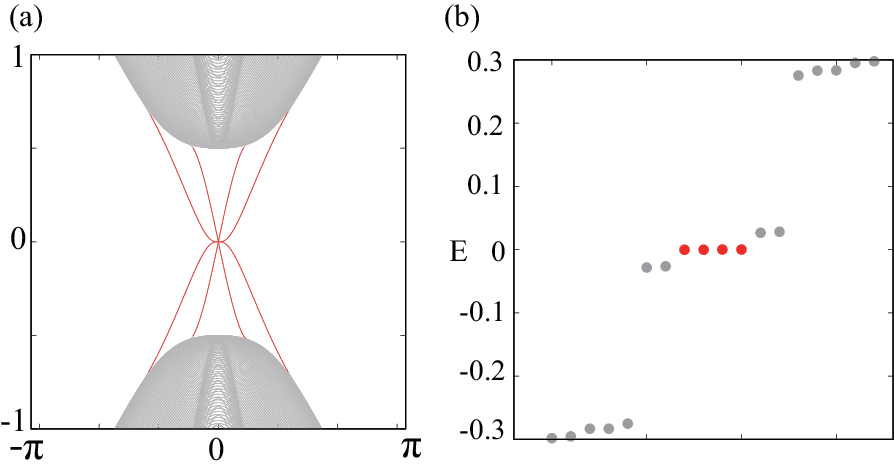}
 \caption{(a) Surface energy spectrum of Eq.~(\ref{eq:doubleTI}) with the parameters $(M,m,t,\delta_x,\delta_y) = (2.5,-1,1.5,0.1,0.1)$ in the ($001$) plane. (b) Eigenvalues (listed in ascending order) of the $s$-wave superconducting states with a vortex line along the $z$ axis, where $\mu=0.1$, $\Delta_0=0.5$, and $\xi=4$ and the lattice sizes $(L_x, L_y, L_z) =(11,11,21)$.}\label{fig:doubleTI}
\end{figure}
\begin{flushleft} 
{\bf S3. C$_4$ symmetry-protected vortex phase transition}
\end{flushleft}

We here demonstrate C$_4$ symmetry-protected Majorana vortex end modes and associated vortex phase transitions. From the analysis of the surface effective Hamiltonian, we see that a simple model having the surface rotation anomaly can be described by a stack of two three-dimensional topological insulators. From this insight, we consider a double topological insulator model: $H(\bm{k})=H_{\rm TI,1} \oplus H_{\rm TI,2}$ with 
\begin{subequations}
\label{eq:TI}
\begin{align}
H_{\rm TI,1}(\bm{k}) = \left(M_1 + m_1 \sum_i \cos (k_i) \right) \bm{1}_2 \otimes \sigma_z + t_1 \sum_i \sin (k_i) s_i \otimes \sigma_x, \\
H_{\rm TI,2}(\bm{k}) = \left(M_2 + m_2 \sum_i \cos (k_i) \right) \bm{1}_2 \otimes \sigma_z + t_2 \sum_i \sin (k_i) s_i \otimes \sigma_x, 
\end{align}
\end{subequations}
where $s_i$ and $\sigma_i$ are the Pauli matrices in the spin and orbital space, respectively. They become a strong topological insulator for $-3 < M_a/m_a < -1$ ($a=1,2$) and host two surface Dirac cones on their surface. 

\begin{figure}[btp]
\centering
 \includegraphics[width=10cm]{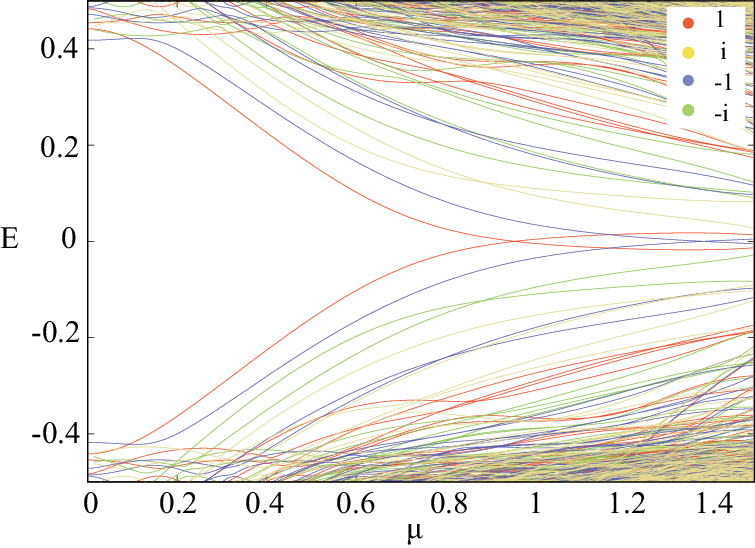}
 \caption{Evolution of vortex bound states as a function of the chemical potential, where  we assume $(M_1,m_1,t_1,M_2,m_2,t_2) = (2.5,-1,1,2.5,-1,1.5)$, $\Delta_0=0.5$, $\xi=4$, and $L_x=L_y =21$. The red, yellow, blue, and green lines represent vortex bound states for C$_4$-subsectors $1$, $i$, $-1$, and $-i$, respectively.}\label{fig:stackTIV}
\end{figure}

Suppose that $s$-wave superconductivity is realized. Then, the BdG Hamiltonian is described as
\begin{align}
\mathcal{H}(\bm{k}) = \begin{pmatrix} H(\bm{k}) - \mu \bm{1}_8 & \Delta_0 \bm{1}_8 \\ \Delta_0^{\ast} \bm{1}_8 & -H(\bm{k}) + \mu \bm{1}_8 \end{pmatrix}, \label{eq:BdG}
\end{align}
where $\mu$ is the chemical potential and $\bm{1}_n$ is the $n \times n$ identity matrix. To see the vortex phase transitions, we implement a vortex line as
\begin{align}
\Delta_0 \to \Delta(\bm{x}) = \Delta_0 \tanh \left( r/\xi \right) e^{i \theta},
\end{align}
where $r = \sqrt{x^2+y^2}$, $\theta  = \arctan (y/x)$, and $\xi$ is the coherence length. The vortex line breaks translation symmetry in the $xy$ plane. For a finite lattice site in the $x$ and $y$ directions, Eq.~(\ref{eq:BdG}) can be described as $\mathcal{H}_{ij}(k_z)$, where $i$ and $j$ indicate lattice sites in the $xy$ plane. Eq.~(\ref{eq:BdG}) with the $U(1)$ vortex field is invariant under C$_4$ symmetry: $\mathcal{C}_4\mathcal{H}(k_z) \mathcal{C}_4^{\dagger} = \tilde{\mathcal{H}}(k_z)$ with
\begin{align}
 \mathcal{C}_4 = e^{-i s_z \frac{\pi}{4}} \otimes \sigma_0 \otimes \mu_{z} \otimes e^{i \tau_z \frac{\pi}{4}}, \label{eq:c4rot}
\end{align}
where $\tilde{\mathcal{H}}(k_z)$ is related to $\mathcal{H}(k_z)$ by C$_4$ rotation of lattice sites in the $xy$ plane; $\mu_i$ and $\tau_i$ are the Pauli matrices in the Nambu space and the stacked degrees of freedom. Numerically diagonalizing the BdG Hamiltonian and Eq.~(\ref{eq:c4rot}), we obtain the evolution of vortex bound states for $k_z=0$ as a function of the chemical potential; see Fig.~\ref{fig:stackTIV}. As expected, we observe $E=0$ level crossings at $\mu_{\rm c,1} \simeq 0.9$ and $\mu_{\rm c,2}\simeq 1.3$, which signal two vortex phase transitions associated with two Majorana vortex zero modes at one end of a vortex line. Each vortex phase transition appears in the subsectors of C$_4$ symmetry with the real eigenvalues $+ 1$ or $-1$. That is to say, the Majorana vortex zero modes and the vortex phase transitions are protected by C$_4$ symmetry.

\end{document}